\documentclass[a4paper]{jpconf}

\usepackage{amsmath,amssymb}
\usepackage{epsfig}
\usepackage[square,sort&compress]{natbib}
\usepackage{graphicx}
\usepackage{float}
\usepackage{array}
\usepackage{rotating}
\usepackage{bm}
\usepackage{url}
\usepackage{color}

\begin{document}
\title{Continuous gravitational wave searches with pulsar 
timing arrays: Maximization versus marginalization over pulsar phase parameters}

\author{Yan Wang$^1$, Soumya D. Mohanty$^2$ and Yi-Qian Qian$^1$}

\address{$^1$ School of Physics, Huazhong University of Science and Technology, 1037 Luoyu Road, Wuhan, Hubei Province 430074, China}

\address{$^2$ Department of Physics, The University of Texas Rio Grande Valley, \\
One West University Blvd, Brownsville, TX 78520, USA}

\ead{ywang12@hust.edu.cn}

\begin{abstract}
Resolvable Supermassive Black Hole Binaries are promising sources
for Pulsar Timing Array based gravitational wave searches.
Search algorithms for such targets 
must contend with the large number of so-called pulsar phase parameters 
in the joint log-likelihood function of the data.
We compare the localization accuracy  for two approaches: 
 Maximization over the pulsar phase parameters (MaxPhase) 
against marginalization over them (AvPhase). Using simulated data from a pulsar timing array 
with 17 pulsars, we find that 
for weak and moderately strong signals, AvPhase  outperforms MaxPhase significantly, while
they perform comparably for strong signals. 
\end{abstract}

\section{Introduction}
Pulsar timing arrays (PTAs) 
aim to detect 
gravitational waves (GWs)   in the 
$10^{-9}-10^{-6}$ Hz range. 
Among the most promising sources in this frequency band are
Supermassive Black Hole Binaries (SMBHBs).
(See \citep{2015RPPh...78l4901L} for a review of the current status of PTA based 
SMBHB searches.) 


The pulsar timing residual induced
by GWs from a non-evolving SMBHB crossing an Earth-pulsar line of sight
depends on the
so-called pulsar phase parameter. This parameter arises from the time-of-flight
of a radio pulse crossing the perturbed space-time between a pulsar and Earth. For a PTA consisting of $N$ pulsars, 
this leads to at least $N$ unknown parameters in the
joint log-likelihood function of all the timing residuals. 


Following the Maximum Likelihood (ML) prescription for 
parameter estimation, the joint log-likelihood function  
must be maximized over all signal parameters, including the $N$ pulsar phases.
Direct numerical maximization over pulsar phases~\citep{2014ApJ...795...96W}
is not scalable when $N$ becomes large. 
A scalable algorithm (MaxPhase),  where the pulsar phases are maximized 
semi-analytically, was proposed in~\citep{2015ApJ...815..125W}. (The code is available from 
\url{https://github.com/yanwang2012/RAAPTR}). This algorithm was applied to a 
prospective PTA in the Square Kilometer Array (SKA) era in \citep{2016arXiv161109440W}. 
While the above algorithms follow the Frequentist canon of maximum likelihood, 
an alternative approach, which is natural in Bayesian statistics, is to
treat pulsar phases as nuisance parameters and marginalize over them. 
Maximizing the marginalized likelihood over the remaining parameters 
provides an {\it ad hoc} Frequentist 
estimator, which we call AvPhase. 
In this paper, we present a brief comparison of MaxPhase 
and AvPhase.

\section{Results and discussion}\label{sec:results}

We simulated a PTA data set with 17 pulsars which are 
observed for 5 years with biweekly cadence. The timing precision for each pulsar 
is set to be 100 ns. The strength of the GW signal is characterized by the network 
signal-noise-ratio $\rho$, which is chosen to be 100, 30 and 8 corresponding to
 strong, moderate and weak signal scenarios. 

\begin{figure}
\centerline{\includegraphics[scale=0.48]{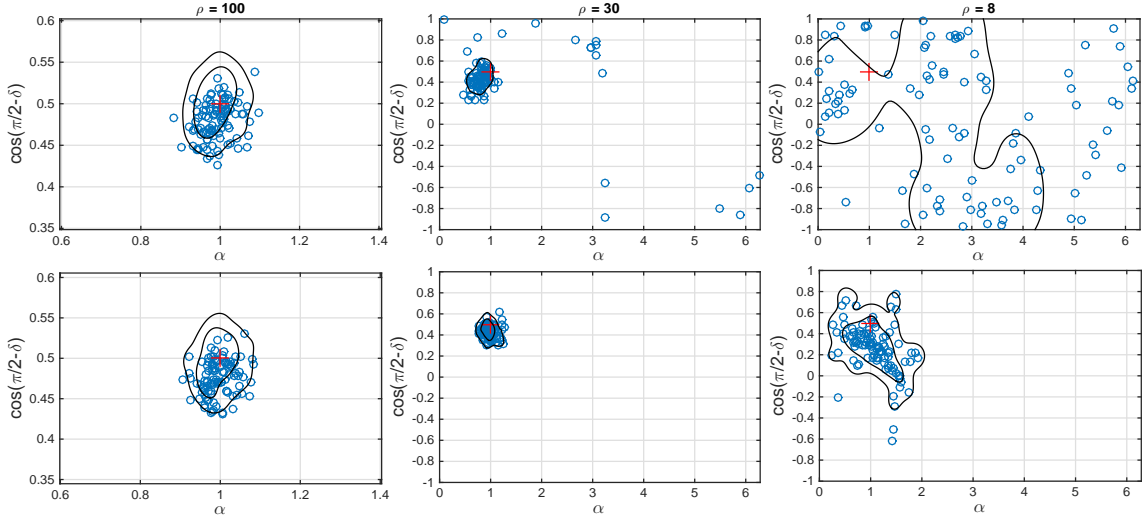}}
\caption{Estimated sky locations (circles) 
obtained from MaxPhase (top row) and avPhase 
(bottom row). 
A red plus sign marks the true location of the source, with
100 independent data realizations used for each location.
The Right Ascension and  Declination of the source are denoted by $\alpha$ and 
$\delta$ respectively. 
The $68\%$ and $95\%$ confidence contours are overlaid.  
}
\label{fig:fig1}
\end{figure}

Fig.~\ref{fig:fig1} shows the estimated sky locations of the GW source 
for the three scenarios. In order of decreasing $\rho$,
the $68\%$ ($95\%$) confidence contours for MaxPhase cover $18~\text{deg}^2$ 
($49~\text{deg}^2$), $354~\text{deg}^2$ (all sky), and $14785~\text{deg}^2$ 
(all sky) respectively; 
while the corresponding ones for avPhase cover  $18~\text{deg}^2$ 
($46~\text{deg}^2$), $111~\text{deg}^2$ ($369~\text{deg}^2$), and $1162~\text{deg}^2$ 
($3535~\text{deg}^2$) respectively. 

Our results show that the localization of avPhase performs 
significantly better than MaxPhase in the weak and moderate signal scenario, 
and its performance is comparable to MaxPhase in the strong signal scenario.  

\ack
Y.W. is supported by the National Natural Science Foundation of China 
under grants 11503007, 91636111 and 11690021. 
The contribution of S.D.M. to this paper is supported by NSF 
awards PHY-1505861 and HRD-0734800. 



\begin{thebibliography}{1}
\expandafter\ifx\csname url\endcsname\relax
  \def\url#1{{\tt #1}}\fi
\expandafter\ifx\csname urlprefix\endcsname\relax\def\urlprefix{URL }\fi
\providecommand{\eprint}[2][]{\url{#2}}

\bibitem{2015RPPh...78l4901L}
{Lommen} A~N 2015 {\em Reports on Progress in Physics\/} {\bf 78} 124901

\bibitem{2014ApJ...795...96W}
{Wang} Y, {Mohanty} S~D and {Jenet} F~A 2014 {\em \apj\/} {\bf 795} 96
  (\textit{Preprint} \eprint{1406.5496})

\bibitem{2015ApJ...815..125W}
{Wang} Y, {Mohanty} S~D and {Jenet} F~A 2015 {\em \apj\/} {\bf 815} 125
  (\textit{Preprint} \eprint{1506.01526})


\bibitem{2016arXiv161109440W}
{Wang} Y, and {Mohanty} S~D 2015 {\em \PRL} {\bf 118} 151104
  (\textit{Preprint} \eprint{1611.09440})

\end{thebibliography}

\providecommand{\newblock}{}

\end{document}